\documentclass[iop]{emulateapj}

\usepackage{epstopdf}
\usepackage{natbib}
\usepackage{hyperref}

\newcommand{\src}{XTE J1701$-$462}
\newcommand{\xte}{{\it RXTE}}
\newcommand{\sw}{{\it Swift}}
\newcommand{\cxo}{{\it Chandra}}
\newcommand{\xmm}{{\it XMM-Newton}}
\newcommand{\pap}{Paper I}

\shorttitle{Variable Quiescent Emission of XTE J1701$-$462}
\shortauthors{Fridriksson et al.}

\journalinfo{The Astrophysical Journal, 736:162, 2011 August 1}

\slugcomment{}

\begin{document}

\title{The Variable Quiescent X-ray Emission of the Neutron~Star~Transient~XTE~J1701$-$462}

\author{Joel~K.~Fridriksson\altaffilmark{1,2}, Jeroen~Homan\altaffilmark{2}, Rudy~Wijnands\altaffilmark{3}, Edward~M.~Cackett\altaffilmark{4}, Diego~Altamirano\altaffilmark{3}, Nathalie~Degenaar\altaffilmark{3}, Edward~F.~Brown\altaffilmark{5}, Mariano~M{\'e}ndez\altaffilmark{6}, and Tomaso~M.~Belloni\altaffilmark{7}}

\altaffiltext{1}{Department of Physics, Massachusetts Institute of Technology, 77 Massachusetts Avenue, Cambridge, MA 02139, USA; joelkf@mit.edu}
\altaffiltext{2}{MIT Kavli Institute for Astrophysics and Space Research, 77 Massachusetts Avenue, Cambridge, MA 02139, USA}
\altaffiltext{3}{Astronomical Institute ``Anton Pannekoek,'' University of Amsterdam, Science Park 904, 1098 XH Amsterdam, The Netherlands}
\altaffiltext{4}{Institute of Astronomy, University of Cambridge, Madingley Road, Cambridge CB3 0HA, UK}
\altaffiltext{5}{Department of Physics and Astronomy, Michigan State University, 3250 Biomedical and Physical Sciences Building, East Lansing, MI 48824, USA}
\altaffiltext{6}{Kapteyn Astronomical Institute, University of Groningen, P.O. Box 800, 9700 AV Groningen, The Netherlands}
\altaffiltext{7}{INAF-Osservatorio Astronomico di Brera, Via E. Bianchi 46, I-23807 Merate (LC), Italy}

\begin{abstract}

We present the results of continued monitoring of the quiescent neutron star low-mass X-ray binary \src\ with \cxo\ and \sw. A new \cxo\ observation from 2010 October extends our tracking of the neutron star surface temperature from $\simeq$800 days to $\simeq$1160 days since the end of an exceptionally luminous 19 month outburst. This observation indicates that the neutron star crust may still be slowly cooling toward thermal equilibrium with the core; another observation further into quiescence is needed to verify this. The shape of the overall cooling curve is consistent with that of a broken power law, although an exponential decay to a constant level cannot be excluded with the present data. To investigate possible low-level activity, we conducted a monitoring campaign of \src\ with \sw\ during 2010 April--October. Short-term flares---presumably arising from episodic low-level accretion---were observed up to a luminosity of $\sim$$1\times10^{35}\textrm{ erg s}^{-1}$, $\sim$20 times higher than the normal quiescent level. We conclude that flares of this magnitude are not likely to have significantly affected the equilibrium temperature of the neutron star and are probably not able to have a measurable impact on the cooling curve. However, it is possible that brighter and longer periods of low-level activity have had an appreciable effect on the equilibrium temperature.

\end{abstract}

\keywords{accretion, accretion disks -- stars: neutron -- X-rays: binaries -- X-rays: individual (XTE J1701$-$462)}

\section{Introduction}\label{sec:intro}
Observations of quiescent neutron star transients have the potential to constrain the internal properties of neutron stars. Some transients (so-called quasi-persistent transients) undergo years- or decades-long outbursts during which the neutron star crust is heated out of thermal equilibrium with the core due to nuclear reactions induced deep in the crust (so-called deep crustal heating; see, e.g., \citealt{brown1998,haensel2008}); observing the cooling of the surface after the source returns to quiescence can give information on the properties of the crust (e.g., \citealt{rutledge2002}). In addition, the equilibrium crust and core temperatures in accreting neutron stars are set by the long-term time-averaged mass accretion rate and the efficiency of the cooling mechanisms (neutrino emission) at work in the neutron star interior; this cooling in turn depends on the properties of the matter inside the star (for an overview on neutron star cooling, see, e.g., \citealt{yakovlev2004}). Measurement of the equilibrium surface temperature can therefore yield information on the material deep inside the neutron star. This, however, also requires good knowledge of the average mass accretion rate onto the star, and this rate is often poorly known. 

\src\ is a transient neutron star low-mass X-ray binary (NS-LMXB) that underwent an exceptionally luminous 19 month outburst in 2006--2007 \citep{lin2009a,homan2010}. We have monitored the source since it entered quiescence in 2007 August; in \citet[][hereafter referred to as \pap]{fridriksson2010} we reported on {\it Rossi X-ray Timing Explorer} (\xte\,) and \sw\ observations tracking the transition of the source from outburst to quiescence, and on \cxo\ and \xmm\ observations covering the first $\simeq$800 days of the quiescent phase. The effective surface temperature of the neutron star---inferred from the thermal component of the quiescent spectra---exhibited a large decrease during the first $\sim$200 days of quiescence, presumably due to the rapid initial cooling of the neutron star crust; such rapid cooling strongly indicates a highly conductive crust \citep{shternin2007,brown2009}. An observation $\simeq$230 days into quiescence showed a large increase in both thermal and nonthermal flux from the source. A subsequent observation ($\sim$70 days later) suggested still somewhat elevated thermal flux; later observations ($\gtrsim$200 days after the initial increase) were consistent with slow cooling from the level preceding the increase. After $\simeq$800 days of quiescence, it was not clear whether the neutron star crust had already reached thermal equilibrium with the core or was still slowly cooling.

\src\ is one of only four NS-LMXBs for which the cooling of the neutron star crust has been monitored after long periods of accretion-induced heating; the other three are KS 1731$-$260 \citep{wijnands2001,wijnands2002a,cackett2006,cackett2010b}, MXB 1659$-$29 \citep{wijnands2003,wijnands2004a,cackett2006,cackett2008}, and EXO 0748$-$676 \citep{degenaar2009b,degenaar2011a}. In addition, a recent observation of the Terzan 5 globular cluster transient IGR J17480$-$2446 indicates that the neutron star crust may have been heated significantly during a very bright $\sim$2 month outburst \citep{degenaar2011b,degenaar2011c}; further observations are needed to confirm this by observing cooling. As discussed in \pap, the quiescent behavior of \src\ has exhibited several noteworthy characteristics. Significantly higher effective surface temperatures have been observed from \src\ than the other three sources; this may be partly due to a high core temperature in the neutron star. The cooling data from the first $\simeq$800 days of quiescence are well fitted---when excluding the observations affected by the temporary increase in temperature---with both an exponential decay to a constant level ($e$-folding time of $\simeq$$120^{+30}_{-20}$ days) and with a broken power-law model. The latter is particularly relevant, since \citet{brown2009} calculate that the general form of the cooling curve should be a broken power law leveling off to a constant at late times. However, the predicted break---mainly due to a change in heat capacity where the crust material transitions from a classical to a quantum crystal---is expected to occur a few hundred days into quiescence, whereas the observed break we reported in \pap\ is at $\sim$20--150 days, and may therefore have a different origin. In addition to the thermal spectral component, \src\ has shown a prominent and highly variable nonthermal component throughout the quiescent phase. Nonthermal components have been seen in the quiescent emission of many NS-LMXBs and are usually well fitted with a simple power law with a photon index of 1--2. The origin of this nonthermal emission is poorly understood; pulsar shock emission or low-level accretion onto the neutron star surface or magnetosphere have been suggested as explanations \citep{campana1998}. In \pap, we speculated that the power-law component seen from \src\ and the large increase in luminosity more than 200 days into quiescence mentioned above are due to episodic low-level residual accretion.

During 2010 April--October we conducted a monitoring program of \src\ using \sw, with short ($\sim$3 ks) observations taking place once every two weeks. The goal was to study possible low-level activity during quiescence---as suggested by the temporary increase in luminosity seen during our \cxo\ and \xmm\ monitoring---and investigate whether such activity can to some extent explain the high surface temperatures observed for \src. In 2010 October, we also obtained a new \cxo\ observation of the source to constrain possible ongoing cooling of the neutron star crust. In this paper, we report on the results of our \sw\ monitoring campaign and recent \cxo\ observation.

\section{Data Reduction and Results}\label{sec:analysis}

\subsection{Cooling Analysis}\label{sec:cxo+xmm}

\src\ was observed with \cxo\ on 2010 October 11--12. This ACIS-S \citep{garmire2003} observation (ObsID 11087) was performed in Timed Exposure mode, with the source located at the default S3 aimpoint. The detector was operated in full-frame mode (with a frame time of 3.2 s), using the Very Faint telemetry format. The observation had a live exposure time of 56.60 ks; no periods of background flaring were found. We analyzed the data with CIAO \citep{fruscione2006}, version 4.2 (CALDB, ver.\ 4.3.1), and with ACIS Extract \citep{broos2010}, version 2010-07-09. At the same time, we also re-analyzed all older \cxo\ observations of the source in quiescence to ensure complete consistency. For information on the earlier observations, see \pap. We mostly follow the same analysis procedure as described in \pap\ and refer to the description therein, mentioning here mainly any differences in the procedure. For each observation, background counts were extracted from an annulus with an inner radius of $\simeq$$5\farcs3$ ($\simeq$10.8 pixels) and an outer radius of $\simeq$$14\farcs3$ ($\simeq$29.0 pixels). All observations were checked to see whether applying a temperature-dependent charge transfer inefficiency correction to the data was appropriate. In all cases, the focal plane temperature stayed within half a degree of the nominal temperature of $-119.7\degr\textrm{C}$ throughout the observation, and therefore no correction was necessary. Two of our 11 \cxo\ observations consisted of two or more exposures spread over a few days. The spectra from the four subexposures of the seventh \cxo\ observation---taken over a period of $\sim$3 days---were combined into a single spectrum using ACIS Extract, as were the response files for the exposures. This was done to facilitate $\chi^2$ fitting of the spectrum; the shortest exposure only had $\simeq$150 counts. As in \pap, the \cxo\ spectra were grouped with the ACIS Extract tool {\tt ae\_group\_spectrum}, with an average of $\sim$25--30 counts per group. The tenth \cxo\ observation---consisting of two exposures performed over a $\simeq$27 hr period---was treated in the same way as the seventh one (note, however, the discussion in \pap\ of possible variability between the individual exposures in both the seventh and tenth observations).

We also re-analyzed the three \xmm\ observations previously reported on in \pap\ with the latest software and calibration (SAS, ver.\ 10.0.2). The analysis proceeded in much the same way as described in our previous paper, and the description of identical analysis steps will not be repeated here. The \xmm\ spectra were grouped with the {\tt specgroup} task, requiring a minimum signal-to-noise ratio of 5, and limiting any oversampling of the intrinsic detector resolutions to less than a factor of 2.5. This resulted in an average number of counts per group of $\sim$35--40, except for the pn spectrum of the third \xmm\ observation, which had an average of $\simeq$51 counts per group.

Using XSPEC \citep{arnaud1996}, version 12.6.0, we fitted all the \cxo\ and \xmm\ spectra simultaneously in the 0.5--10 keV band, except for the third \xmm\ observation (XMM-3), which we fitted separately. As discussed in \pap, this spectrum is very different from the others, showing much higher nonthermal flux and a significant increase in thermal flux compared to the previous observation; XMM-3 will be discussed further below. We used the {\tt nsatmos} neutron star atmosphere model (\citealt{heinke2006}; see also a discussion of this model in \pap) plus a simple power-law model ({\tt pegpwrlw}), modified by photoelectric absorption. We used the {\tt TBnew} absorption model\footnote{See \url{http://pulsar.sternwarte.uni-erlangen.de/wilms/research/tbabs/}.}, an updated version of the {\tt TBabs} model \citep{wilms2000}, with the {\tt vern} cross sections \citep{verner1996} and {\tt wilm} abundances \citep{wilms2000}. All the {\tt TBnew} parameters except the equivalent hydrogen column density were kept fixed at their default values. The {\tt TBabs}/{\tt TBnew} model improves in various ways on older absorption models \citep{wilms2000} and is therefore to be preferred. In the {\tt nsatmos} model, we fixed the neutron star mass at a value of 1.4 $M_\sun$, the distance at 8.8 kpc \citep{lin2009b}, and the fraction of the neutron star surface emitting at 1. As in \pap, we tied the absorption column and photon index between all observations (except for XMM-3). Although we have no reason to expect the photon index to necessarily have the same value for all the observations, our spectra do not have enough counts to allow it to vary freely for each observation (see further discussion of this and its possible effects on our results in \pap). We allowed the neutron star radius to float to its best-fit value initially (10.5 km), and then fixed it at this value when performing error scans for other free parameters (an error scan for the radius gives an uncertainty of $\pm$2.5 km).

\begin{deluxetable*}{lrccccc}
\tablewidth{0pt}
\tabletypesize{\footnotesize}
\tablecaption{Derived Parameter Values from {\it Chandra} and {\it XMM-Newton} Observations\label{results}}
\tablehead{\colhead{Observation\tablenotemark{a}} & \colhead{$t-t_0$\tablenotemark{b}} & \colhead{$kT^{\infty}_\mathrm{eff}$\tablenotemark{c}} & \colhead{$F_\mathrm{pl}$\tablenotemark{d}} & \colhead{PL Fraction\tablenotemark{e}} & \colhead{$L_\mathrm{bol}$\tablenotemark{f}} & \colhead{$L_\mathrm{tot}$\tablenotemark{g}}\\
\colhead{} & \colhead{(days)} & \colhead{(eV)} & \colhead{($10^{-13}\textrm{ erg s}^{-1}\textrm{ cm}^{-2}$)} & \colhead{(\%)} & \colhead{($10^{33}\textrm{ erg s}^{-1}$)} & \colhead{($10^{33}\textrm{ erg s}^{-1}$)}
}
\startdata
CXO-1  & 2.95 & $163.1\pm3.5$ & $\phn1.5\pm1.1$ & $\phn9\pm6$ & $16.6\pm1.5$ & $15.7\pm1.7$\\
CXO-2  & 10.81 & $158.2\pm2.5$ &  $\phn1.4\pm0.7$ & $\phn9\pm4$ & $14.7\pm1.0$ & $13.8\pm1.2$\\
XMM-1 & 16.24 & $155.2\pm1.3$ & $\phn0.5\pm0.3$ & $\phn4\pm2$ & $13.6\pm0.5$ & $12.0\pm0.5$\\
XMM-2 & 49.49 & $149.2\pm1.3$ & $\phn0.8\pm0.3$ & $\phn7\pm2$ & $11.6\pm0.4$ & $10.5\pm0.5$\\
CXO-3  & 174.33 & $128.7\pm4.7$ & $\phn4.9\pm0.9$ & $47\pm6$ & $\phn6.4\pm0.9$ & $\phn9.6\pm0.7$\\
XMM-3 & 225.72 & $157.8\pm1.8$ & $15.2\pm0.6$ & $53\pm2$ & $14.6\pm0.7$ & $26.5\pm0.5$\\
CXO-4  & 298.30 & $135.1\pm2.0$ & $\phn1.4\pm0.3$ & $17\pm3$ & $\phn7.8\pm0.5$ & $\phn7.6\pm0.6$\\
CXO-5  & 431.07 & $125.5\pm3.1$ & $\phn3.1\pm0.5$ & $39\pm5$ & $\phn5.8\pm0.6$ & $\phn7.4\pm0.5$\\
CXO-6  & 540.08 & $125.0\pm1.5$ & $\phn0.5\pm0.1$ & $10\pm2$ & $\phn5.7\pm0.3$ & $\phn5.0\pm0.3$\\
CXO-7  & 592.68 & $128.4\pm2.3$ & $\phn2.2\pm0.4$ & $29\pm4$ & $\phn6.4\pm0.5$ & $\phn7.1\pm0.4$\\
CXO-8  & 652.62 & $123.3\pm2.2$ & $\phn1.8\pm0.3$ & $28\pm4$ & $\phn5.4\pm0.4$ & $\phn5.9\pm0.3$\\
CXO-9  & 705.38 & $123.0\pm2.0$ & $\phn1.5\pm0.3$ & $25\pm3$ & $\phn5.4\pm0.3$ & $\phn5.6\pm0.3$\\
CXO-10 & 795.63 & $123.7\pm1.7$ & $\phn1.1\pm0.2$ & $19\pm3$ & $\phn5.5\pm0.3$ & $\phn5.3\pm0.3$\\
CXO-11 & 1158.84 & $120.6\pm1.7$ & $\phn0.8\pm0.2$ & $16\pm3$ & $\phn5.0\pm0.3$ & $\phn4.6\pm0.3$
\enddata
\tablenotetext{a}{See \cite{fridriksson2010} for technical information on all observations except the latest one.}
\tablenotetext{b}{Time of mid observation; $t_0$ is MJD 54321.95.}
\tablenotetext{c}{Effective surface temperature of the neutron star as seen by an observer at infinity.}
\tablenotetext{d}{Unabsorbed 0.5--10 keV flux of the power-law component.}
\tablenotetext{e}{Fractional contribution of the power-law component to the total unabsorbed 0.5--10 keV flux.}
\tablenotetext{f}{Unabsorbed bolometric luminosity of the thermal ({\tt nsatmos}) component.}
\tablenotetext{g}{Unabsorbed total luminosity in the 0.5--10 keV band.}
\tablecomments{All numbers were derived assuming a neutron star radius of 10.53 km, a mass of 1.4 $M_\sun$, and a distance of 8.8 kpc. Errors quoted are at the $1\sigma$ Gaussian (68.3\%) confidence level.}
\end{deluxetable*}

\tabletypesize{\scriptsize}
\begin{deluxetable*}{ccccccccccc}
\tablewidth{0pt}
\tablecaption{Best-fit Cooling Curve Parameters\label{cooling_param}}
\tablehead{\multicolumn{4}{c}{Exponential Decay Fit} & \colhead{} & \multicolumn{5}{c}{Broken Power-law Fit} & \colhead{Excluded\tablenotemark{i}} \\
\cline{1-4} \cline{6-10}
\colhead{$\tau$\tablenotemark{a}}  & \colhead{$kT_\mathrm{eq}$\tablenotemark{b}} & \colhead{$kT'$\tablenotemark{c}} & \colhead{$\chi^2_\nu$ (dof)\tablenotemark{d}} & \colhead{} & \colhead{$\gamma_1$\tablenotemark{e}} & \colhead{$\gamma_2$\tablenotemark{f}} & \colhead{$t_\mathrm{b}$\tablenotemark{g}} & \colhead{$A$\tablenotemark{h}} & \colhead{$\chi^2_\nu$ (dof)\tablenotemark{d}} & \\
\colhead{(days)}  & \colhead{(eV)} & \colhead{(eV)} & \colhead{} & \colhead{} & \colhead{} & \colhead{} & \colhead{(days)} & \colhead{(eV)} & \colhead{} & \colhead{}}
\startdata
$133_{-25}^{+38}$ & $123.4\pm0.9$ & $36.9\pm1.7$ & 1.07 (9) && $0.030\pm0.013$ & $0.069\pm0.004$ & $38_{-12}^{+24}$ & $168.8\pm5.7$ & 0.88 (8) & XMM-3 and CXO-4\\
$230\pm46$ & $121.9\pm1.5$ & $35.8\pm1.4$ & \phn1.66 (10) && $0.030\pm0.012$ & $0.069\pm0.004$ & $40_{-13}^{+42}$ & $168.8\pm5.5$ & 1.13 (9) & XMM-3
\enddata
\tablenotetext{a}{Best-fit $e$-folding time of the decay.}
\tablenotetext{b}{Best-fit constant offset to the decay.}
\tablenotetext{c}{Best-fit decay amplitude.}
\tablenotetext{d}{Reduced $\chi^2$ for the fit and number of degrees of freedom.}
\tablenotetext{e}{Best-fit pre-break power-law slope.}
\tablenotetext{f}{Best-fit post-break power-law slope.}
\tablenotetext{g}{Best-fit break time between power laws.}
\tablenotetext{h}{Best-fit pre-break normalization coefficient (i.e., temperature at $t-t_0=1$ day).}
\tablenotetext{i}{Observations excluded from the fit.}
\tablecomments{Errors quoted are at the $1\sigma$ Gaussian (68.3\%) confidence level.}
\end{deluxetable*}
\tabletypesize{\footnotesize}

\begin{figure}[t]
\centerline{\includegraphics[width=8.4cm,trim=5 60 55 70,clip=true]{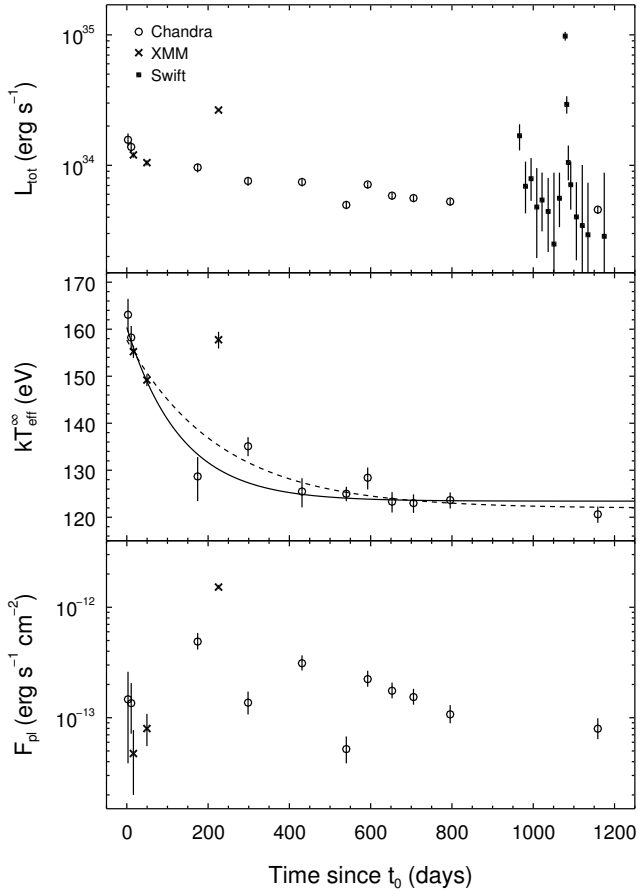}}
\caption{Spectral evolution of \src\ since the end of the 2006--2007 outburst. Top panel: unabsorbed luminosity in the 0.5--10 keV band. We do not show a $3\sigma$ \sw\ upper limit to the luminosity (observation 16; shown in Figure~\ref{fig:lc_swift}), which is virtually coincident with the last \cxo\ data point. Middle panel: effective neutron star surface temperature. The solid line is the best-fit exponential decay to a constant level, excluding the sixth and seventh data points. The dashed line is a similar fit including the seventh data point. Bottom panel: unabsorbed 0.5--10 keV power-law flux.}\label{fig:lum+cooling+pl}
\end{figure}

\begin{figure}[t]
\centerline{\includegraphics[width=8.6cm,trim=10 10 50 35,clip=true]{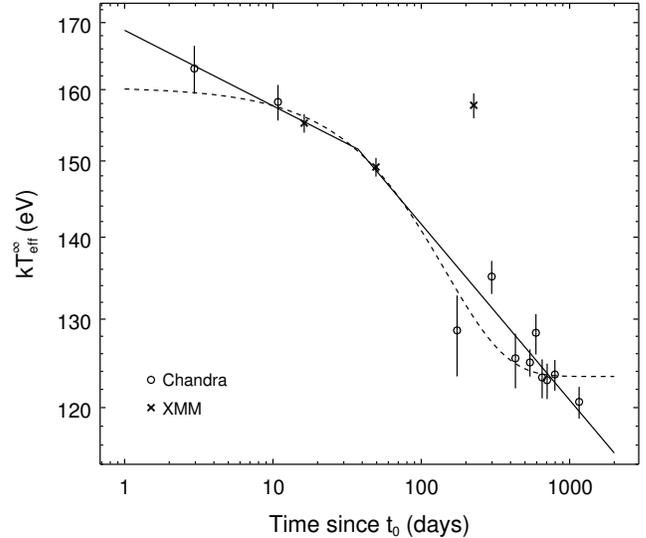}}
\caption{Effective neutron star surface temperature during quiescence, with best-fit cooling curves shown. The solid curve is a broken power law fitted to all data points except the sixth and seventh ones; including data point 7 in the fit yields a virtually identical curve. The dashed curve is the best-fit exponential decay curve with a constant offset, where points 6 and 7 have been excluded from the fit (shown as a solid curve in Figure~\ref{fig:lum+cooling+pl}).}\label{fig:cooling_log}
\end{figure}

In Table~\ref{results}, we show the (gravitationally redshifted) effective neutron star surface temperature for each individual observation, as inferred from our spectral fits. Uncertainties in the neutron star radius, mass, and distance give rise to a systematic uncertainty in the temperatures. This uncertainty is highly correlated between the different observations, and we do not include it in the error bars in Figures~\ref{fig:lum+cooling+pl} and \ref{fig:cooling_log} and Tables~\ref{results} and \ref{cooling_param}. Changes in the radius, mass, or distance systematically shift all the temperature values in more or less the same way (the higher the temperature, the larger the shift); in general, our conclusions about the cooling would not be significantly affected by such shifts. To assess the magnitude of this uncertainty, we changed the values of these three parameters one at a time and re-fitted the spectra (allowing the absorption column and photon indices to float to new best-fit values). We consider a neutron star radius range of 8--15 km, a mass range of 1.1--2.0 $M_\sun$, and a distance range of 7.5--10.1 kpc (based on the 1.3 kpc uncertainty in the distance estimate; \citealt{lin2009b}). Changing the radius to 8 km (15 km) results in increases in the temperatures of $\simeq$6.2--10.0 eV (decreases of $\simeq$9.0--13.5 eV). Changing the mass to 1.1 $M_\sun$ (2.0 $M_\sun$) gives temperature increases of $\simeq$2.5--4.1 eV (decreases of $\simeq$5.5--8.6 eV). Finally, changing the distance to 7.5 kpc (10.1 kpc) yields decreases of $\simeq$6.3--9.2 eV (increases of $\simeq$5.7--8.7 eV). Overall, we estimate that this gives rise to a combined systematic uncertainty of $\sim$10 eV in the temperatures measured in recent observations. We stress that the parameter values derived from our cooling curve fits (see below) are not significantly affected by this (with the exception of the baseline temperature of the exponential fit). This is further discussed and quantified in \pap; there we also discuss uncertainty arising from the handling of the nonthermal component. We note here that all errors quoted in this paper correspond to $1\sigma$ Gaussian (68.3\%) confidence.

In Table~\ref{results}, we also show the 0.5--10 keV power-law flux and total luminosity, and the bolometric luminosity (all unabsorbed), as well as the fractional contribution of the power-law component to the total unabsorbed 0.5--10 keV flux. We plot the luminosity, temperature, and power-law flux in Figure~\ref{fig:lum+cooling+pl}. The best-fit values for the equivalent hydrogen column density and the tied photon index are $(2.98\pm0.04)\times10^{22}\textrm{ cm}^{-2}$ and $1.96\pm0.20$. The quality of the simultaneous fit is good: the reduced $\chi^2$ ($\chi^2_\nu$) is 0.94 for 314 degrees of freedom (dof). The separate fit to XMM-3 is worse: $\chi^2_\nu=1.30$ (197 dof). In particular, this fit tends to underestimate the flux in the lowest energy bins, and the spectrum prefers a lower value for the absorption column. However, it is rather implausible that this observation would have significantly lower absorption than other observations during both quiescence and outburst (see below), and we therefore fix the column density at the best-fit value from the simultaneous fit (quoted above). The best-fit value of the XMM-3 photon index is $1.40\pm0.08$. We also tried fitting the XMM-3 spectrum with the power-law component replaced by the {\tt simpl} or {\tt compTT} Comptonization models, but we were unable to improve on the fit. We note that it is conceivable that this spectrum is affected by the presence of elements heavier than hydrogen in the neutron star atmosphere. For an accretion rate above $\dot{M}_Z\sim4\times10^{-14}\textrm{ }M_\sun\textrm{ yr}^{-1}\textrm{ }(8/Z)(kT_\mathrm{eff}/0.1\textrm{ keV})^{3/2}$, an element with atomic number $Z$ is not expected to settle quickly enough out of the atmosphere for it to maintain a pure hydrogen composition \citep{brown1998,bildsten1992}. The observed luminosity in XMM-3 corresponds to $\dot{M}\sim2\times10^{-12}\textrm{ }M_\sun\textrm{ yr}^{-1}$ (assuming $L=\epsilon\dot{M}c^2$, with $\epsilon=0.2$, and that $\sim$80\% of the emission arises from accretion), implying that a presence of helium is possible if most or all of the additional flux---compared to the thermal emission of the previous observation---stems from accretion onto the neutron star surface. (We can also not exclude the possibility that some of the other observations are affected by heavier elements in the atmosphere, although this is less likely; the fact that those spectra are in general well fitted by a pure hydrogen atmosphere model indicates that this is probably not a large effect if present.) Another possible reason for the low quality of the XMM-3 fit is that thermal flux arising from ongoing accretion during this observation acts to distort the shape of the thermal component from that of a neutron star atmosphere spectrum. Finally, we note that it is conceivable that cross-calibration errors play a role here. \cite{tsujimoto2011} explore the cross-calibration between various X-ray detectors and derive lower absorption column values for \xmm\ EPIC data, especially the pn detector, than for \cxo\ ACIS data. For XMM-3, the pn spectrum contributes the majority of the residuals. However, even when allowing the column density to be free in the XMM-3 fit, the quality of the fit is still marginal ($\chi^2_\nu=1.15 $ for 196 dof), and the change in the column density (lower by $\simeq$14\%) is somewhat larger than seen by \cite{tsujimoto2011} ($\simeq$11\% for the pn and $\sim$4\%--6\% for the MOS detectors). In addition, the other two \xmm\ observations agree well with the simultaneous fit and show no indication of the pn spectrum preferring a lower column density value than the MOS spectra.

The temperature values reported here for the older \cxo\ and \xmm\ observations are $\simeq$0.4--1.5 eV lower than the corresponding values reported in \pap. This is due to small differences in the data analysis (value used for the neutron star radius; absorption model used; handling of XMM-3, CXO-7, and CXO-10; more recent calibration) and the effect of the new observation on the simultaneous fit. The best-fit column density value quoted above is significantly higher than the one reported in \pap, primarily due to the different abundance table used here. We note that when fitting with the {\tt phabs} absorption model with {\tt angr} abundances and {\tt bcmc} cross sections, as we did in \pap, we get an absorption column of $2.0\times10^{22}\textrm{ cm}^{-2}$, the same as the value derived by \cite{lin2009a} from \xte\ and \sw\ observations during the 2006--2007 outburst (and very close to the value derived in \pap).

The new \cxo\ observation---made $\simeq$1160 days into quiescence---has an effective neutron star surface temperature (as seen by an observer at infinity) $kT_\mathrm{eff}^{\infty}=120.6\pm1.7$ eV, compared to $123.7\pm1.7$ eV in the previous \cxo\ observation a year earlier. These two temperature values differ at the $\simeq$$1.3\sigma$ (i.e., $\sim$80\%) confidence level; using the weighted temperature average for the three \cxo\ observations preceding the latest one, $123.4\pm1.1$ eV, gives similar significance for a change in the temperature. The data are therefore more consistent with continued cooling than with the temperature having reached a fixed equilibrium value, but given the size of the error bars it is not possible to draw a firm conclusion. We also note that uncertainties associated with the handling of the nonthermal component can possibly give rise to small shifts in the temperatures (see discussion in \pap). The current temperature corresponds to a bolometric thermal luminosity of $\simeq$$5.0\times10^{33}\textrm{ erg s}^{-1}$. The power-law flux of the new observation is among the lowest seen during quiescence and represents $\simeq$16\% of the total 0.5--10 keV unabsorbed flux. The 0.5--10 keV unabsorbed luminosity of the new observation ($\simeq$$4.6\times10^{33}\textrm{ erg s}^{-1}$) is the lowest observed so far from the source.

To estimate the end time of the outburst, $t_0$, we followed the same procedure as in \pap. We fitted the luminosity measured in the three \sw\ observations made during the final decay of the outburst with a simple exponential decay function, and did the same for the first three \cxo\ and \xmm\ observations in quiescence; we then define $t_0$ as the intersection of the two curves. We did not re-extract the \sw\ spectra from \pap\ but did re-fit them. The first two spectra were grouped with a minimum of 25 counts per bin and fitted with a simple absorbed power law ({\tt TBnew*pegpwrlw}), fixing the absorption column at the value derived from our observations in quiescence (quoted above). The third spectrum only had 52 counts, and we therefore fitted the unbinned spectrum with the $W$-statistic in XSPEC. In this case, we added an {\tt nsatmos} component to the spectral model, since the luminosity of this observation is low enough for thermal radiation from the neutron star surface to make a significant contribution to the flux (we derive a 0.5--10 keV unabsorbed luminosity of $(3.8\pm0.7)\times10^{34}\textrm{ erg s}^{-1}$). We fixed the effective neutron star surface temperature at a value of $\log(T_\mathrm{eff}/K)=6.4$ (i.e., a gravitationally redshifted value of $kT_\mathrm{eff}^\infty\approx166$ eV, slightly higher than the value measured in the first quiescent observation) and the radius at 10.5 km (as derived from our quiescent spectra). Due to the low number of counts in the spectrum of the third observation, we also in that case fixed the photon index at a value of 2 (similar to the value measured in the previous \sw\ observation and the combined value in our quiescent fit). The resulting value for $t_0$ is MJD 54321.95, i.e., $\simeq$1.4 days after the final \sw\ observation and $\simeq$3.0 days before the first \cxo\ observation. This value for $t_0$ is 0.18 days earlier than the value derived in \pap. The difference arises mostly from a somewhat lower luminosity value derived here for the last of the three \sw\ observations; in \pap\ we did not include an {\tt nsatmos} component in the model for that spectrum.

As in \pap, we fitted our derived temperatures with two models: an exponential decay function with a constant offset, $T_\mathrm{eff}^\infty(t)=T'\exp[-(t-t_0)/\tau]+T_\mathrm{eq}$, and a broken power-law model (as a function of $t-t_0$). In both cases, we kept $t_0$ fixed at the value quoted above. The fits were performed with Sherpa \citep{freeman2001}; errors were estimated using the confidence method.\footnote{See documentation at the Sherpa Web site: \url{http://cxc.cfa.harvard.edu/sherpa/}.} In all our fits we exclude XMM-3; as mentioned above, this observation has a much higher luminosity (arising from increases in both thermal and nonthermal flux) than the preceding and subsequent observations, likely due to an accretion event. We perform fits both with and without the subsequent observation, CXO-4. Although this observation shows some evidence of increased surface temperature, it has neither anomalously high luminosity nor nonthermal flux. It is not clear whether this possibly elevated temperature is somehow due to increased accretion (before or during the observation), nor is it clear whether it is related to the elevated luminosity observed in XMM-3. It may possibly be a statistical fluctuation, or it may be due to our limited ability to separate the contributions of the two spectral components (see further discussion in \pap\ as well as the discussion of possible effects from low-level accretion in Section~\ref{sec:cooling_effects}). The results of our cooling fits are shown in Table~\ref{cooling_param} and best-fit cooling curves are plotted in Figures~\ref{fig:lum+cooling+pl} and \ref{fig:cooling_log}. We note that the value used for $t_0$ cannot affect the $e$-folding time or the equilibrium temperature of the exponential fits. It does affect the decay amplitude, but within the small uncertainty in $t_0$ that effect is negligible (of the order of half an eV). However, the break time, and especially the pre-break slope, of the broken power-law fits are quite sensitive to the value of $t_0$; the post-break slope is practically unchanged within the allowed range for $t_0$. We performed broken power-law fits (both XMM-3 and CXO-4 excluded) with $t_0$ shifted within the allowed range set by the last/first observation in outburst/quiescence. Shifting $t_0$ to two days later (one day earlier) gives best-fit break times of $\simeq$30 ($\simeq$42) days and pre-break slopes of $\simeq$0.019 ($\simeq$0.034). The effects of changes in $t_0$ on broken power-law fits are further discussed in \pap.


Both the exponential and the broken power-law model give acceptable fits to the temperatures when both XMM-3 and CXO-4 are excluded: $\chi^2_\nu= 1.07$ (9 dof) for the exponential and $\chi^2_\nu= 0.88$ (8 dof) for the broken power law. However, an ongoing decrease in the temperature is inconsistent with the exponential fit. When CXO-4 is included, the broken power-law fit is superior: $\chi^2_\nu= 1.13$ (9 dof), compared to $\chi^2_\nu= 1.66$ (10 dof) for the exponential fit. The exponential fit with CXO-4 included (dashed line in Figure~\ref{fig:lum+cooling+pl}) matches the later data points ($t-t_0\gtrsim250$ days) quite well but fits the first five data points badly. In contrast to the broken power-law fits, a single unbroken power law gives poor fits, with $\chi^2_\nu=2.07$ (10 dof) when both XMM-3 and CXO-4 are excluded; the best-fit slope is $0.059\pm0.002$. We note that we cannot exclude the possibility that our inferred temperatures (and thereby the cooling curve fits) are affected by residual accretion; we discuss this in Section~\ref{sec:cooling_effects} and argue that the effects of this, if present, are likely small.


\begin{deluxetable}{llcc}
\tablewidth{0pt}
\tablecaption{\sw\ XRT Observations of \src\ in Quiescence\label{obs_info}}
\tablehead{\colhead{No.} & \colhead{Start Date\tablenotemark{a}} & \colhead{ObsID}  & \colhead{Exp.\ Time\tablenotemark{b} (ks)}}
\startdata
1 & Apr 2 & 00090523001 & 2.90\\
2 & Apr 16 & 00090523002 & 3.43\\
3 & Apr 30 & 00090523003 & 3.36\\
4 & May 14 & 00090523004 & 3.06\\
5 & May 27 & 00090523005 & 2.81\\
6 & Jun 11 & 00090523006 & 2.85\\
7 & Jun 25 & 00090523007 & 2.79\\
8 & Jul 9 & 00090523008 & 3.05\\
9 & Jul 23 & 00090523009 & 3.20\\
10 & Jul 27 & 00031776001 & 3.00\\
11 & Jul 31 & 00031776002 & 3.16\\
12 & Aug 6 & 00090523010 & 2.66\\
13 & Aug 20 & 00090523011 & 3.26\\
14 & Sep 3 & 00090523012 & 1.56\\
15 & Sep 17 & 00090523013 & 3.09\\
16 & Oct 1 & 00090523014 & 3.08\\
17 & Oct 27 & 00090523015 & 2.11
\enddata
\tablenotetext{a}{All dates are in 2010.}
\tablenotetext{b}{Good live exposure time.}
\end{deluxetable}

\subsection{\textit{Swift} Monitoring Program}\label{sec:swift}

\src\ was observed 17 times with the \sw\ X-ray Telescope (XRT; \citealt{burrows2005}) during 2010 April--October (see Table~\ref{obs_info} for details on each observation). Fifteen of these were regularly scheduled observations made approximately every two weeks as part of our monitoring program of the source; two additional Director's Discretionary Time (DDT) observations (nos.\ 10 and 11 in Table~\ref{obs_info}) were made in late July. All the observations were performed with the detector in Photon Counting mode. We analyzed the data with HEASOFT, version 6.10. Using the {\tt xrtpipeline} task, we processed and screened (with the default settings) the Level 1 raw event files and created vignetting-corrected exposure maps for each observation. We then extracted counts (and spectra) in the 0.5--10 keV range from the Level 2 screened event files using Xselect; source counts were extracted from a circle of radius $30\arcsec$ and background counts were in each case extracted from a nearby circle of radius $250\arcsec$. For each observation, we used XIMAGE to integrate the exposure map over both the source and background extraction circles, and used the ratio of the integrated exposures as our background scaling factor, thereby taking into account the effects of vignetting and CCD bad pixels and hot columns. We also used the {\tt xrtmkarf} task (with an exposure map) to calculate a correction factor for each observation to correct the net (i.e., background-subtracted) count rate for losses due to the finite size of the extraction region compared to the point-spread function (PSF), as well as for losses due to vignetting and CCD bad pixels and hot columns. The count rates we quote therefore correspond to an on-axis extraction region of infinite size.

\begin{figure}[t]
\centerline{\includegraphics[width=8.8cm,trim=0 30 45 40,clip=true]{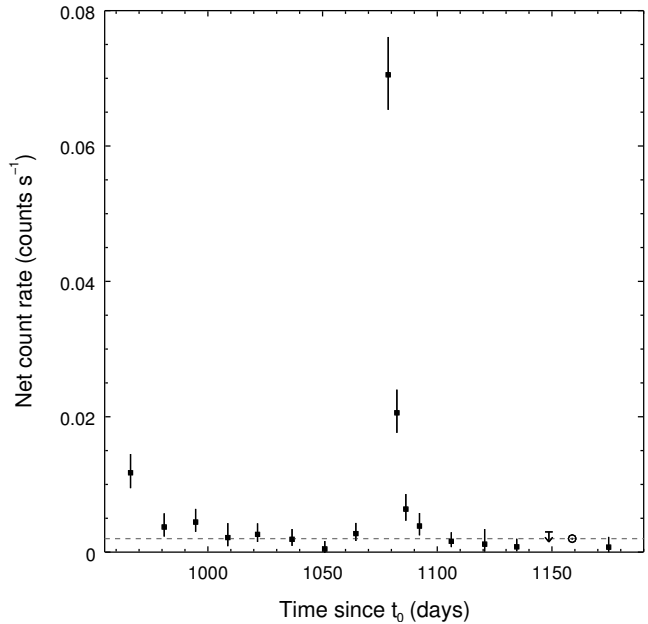}}
\caption{\sw\ XRT light curve of \src\ during 2010 April--October (filled squares); the arrow represents a $3\sigma$ upper limit. The open circle indicates a contemporaneous \cxo\ observation whose corresponding \sw\ XRT count rate was calculated from its spectrum; this count rate level is also indicated by the dashed line.}\label{fig:lc_swift}
\end{figure}

In Figure~\ref{fig:lc_swift}, we show the \sw\ light curve of \src. For one observation, in which the source was not detected, we show a $3\sigma$ upper limit. We also show a data point representing the latest \cxo\ observation (CXO-11); we used XSPEC to calculate the \sw\ XRT count rate corresponding to the best-fit \cxo\ spectrum. This CXO-11 count rate level is also indicated with a gray dashed line; we note that the spectrum from the previous \cxo\ observation---which was performed around 800 days into quiescence and had slightly higher thermal and nonthermal flux---corresponds to a $\sim$40\% higher count rate. Most of the \sw\ observations show count rates consistent with the quiescent level observed in CXO-11. However, the first \sw\ observation shows a clearly elevated count rate (a factor of $\sim$6 increase compared to the expected quiescent level) and the ninth observation shows a much stronger (factor of $\sim$35) increase. The two following DDT observations, performed $\sim$4 and $\sim$8 days later, show a rapidly decaying flux. An observation performed two weeks after the brightest (i.e., ninth) observation shows a weak indication of an elevated count rate, but is consistent with the baseline quiescent level.

\begin{figure*}[t]
\centerline{\includegraphics[width=15cm,trim=0 30 40 50,clip=true]{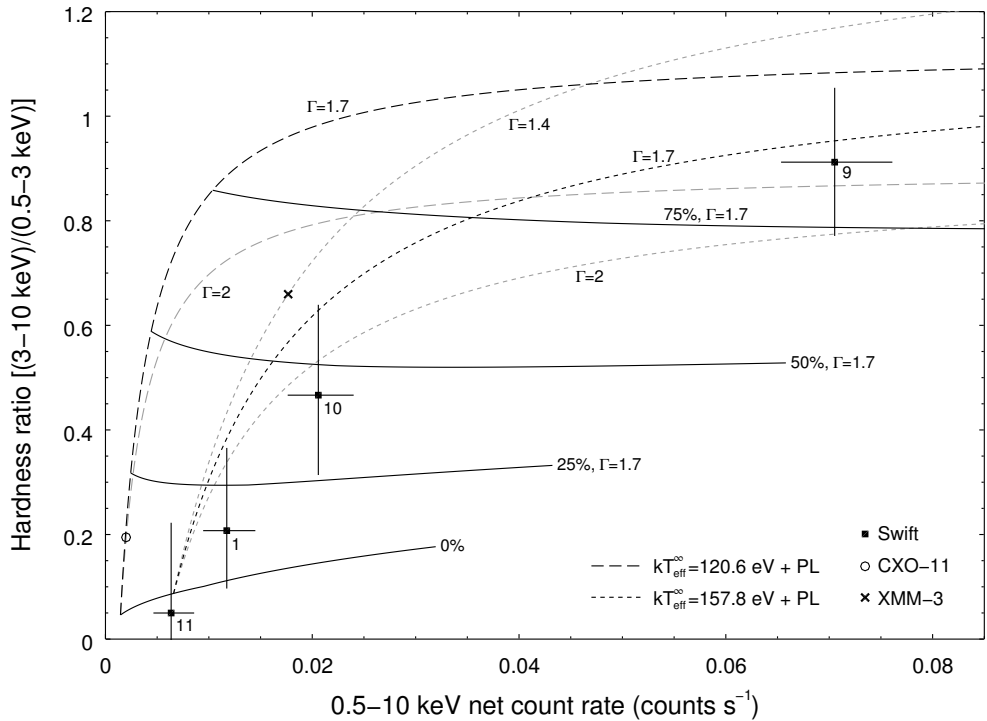}}
\caption{Hardness--intensity diagram for the four brightest \sw\ observations (number of observation indicated next to data point). Also shown are the expected locations of the CXO-11 and XMM-3 spectra, as well as curves corresponding to spectra with various combinations of thermal ({\tt nsatmos}) and nonthermal (power-law) contributions. Along the solid curves, the fractional contribution (indicated next to curve) of the power-law component to the 0.5--10 keV unabsorbed flux is constant while both spectral components vary (a photon index $\Gamma=1.7$ is assumed). Along the dashed curves, the thermal flux is constant (with two different temperature values considered), but the power-law flux varies (the assumed value of the photon index is indicated next to each curve). Further details are given in the text.}\label{fig:hardness}
\end{figure*}

For the three brightest \sw\ observations (nos.\ 1, 9, and 10), we also performed spectral fitting (using XSPEC) in the 0.5--10 keV band. We created ancillary response files using the {\tt xrtmkarf} task and used a standard redistribution matrix file from the calibration database. We used the corrected background scaling factor described above. For observations 1 and 10, which only had 28 and 50 counts, respectively, we fitted the unbinned spectra with the $W$-statistic in XSPEC. We binned the spectrum for observation 9 into groups with a minimum of 26 counts. We used the same spectral model as for the \cxo\ and \xmm\ spectra, fixing all the parameters except the power-law normalization (but including the effective surface temperature) at the best-fit values for CXO-11 as derived from the simultaneous \cxo\ and \xmm\ fit described above. For the brightest \sw\ observation (no.\ 9), we also allowed the photon index to vary; this gave a value of $2.11\pm0.19$, similar to the combined value from our \cxo\ and \xmm\ fit. For all other \sw\ observations (i.e., except the three brightest ones), we estimate a luminosity from the observed count rate, assuming a fixed thermal component equal to that seen in CXO-11 and a varying power-law component with a photon index value equal to that obtained from our simultaneous fit (i.e., we subtract the count rate corresponding to the absorbed CXO-11 thermal component---calculated with XSPEC---from the observed count rate and attribute the rest to an absorbed power-law component with photon index 1.96; we then convert this power-law count rate to an unabsorbed flux and add to the CXO-11 unabsorbed thermal flux). We show the \sw\ luminosities in Figure~\ref{fig:lum+cooling+pl}. The peak luminosity observed in the 2010 July flare is $\simeq$$1.0\times10^{35}\textrm{ erg s}^{-1}$, i.e., a factor of $\sim$20 higher than the expected quiescent level (as seen in our two most recent \cxo\ observations). The luminosity of \sw\ observations 9--13 (i.e., the initial peak detection of the flare and the four subsequent observations) can be fitted ($\chi^2_\nu=0.13$ for 3 dof) with an exponential decay down to a constant level of $\simeq$$4.6\times10^{33}\textrm{ erg s}^{-1}$, equal to that in the latest \cxo\ observation (which took place roughly 2 months later); the best-fit $e$-folding time is $\simeq$2.8 days. 

The limited number of counts in our \sw\ spectra make it difficult to place constraints on the spectral composition and determine to what extent the increases in luminosity are due to the thermal/nonthermal component. Performing an error scan with the temperature free (in addition to the power-law normalization and slope) for the spectrum of the brightest \sw\ observation (no.\ 9, with $\simeq$180 counts) yields a 90\% confidence upper limit on the effective temperature of $kT_\mathrm{eff}^\infty\simeq199$ eV (and no lower limit). Fixing the temperature at that value and re-fitting the spectrum results in a fractional contribution of the thermal flux to the total unabsorbed 0.5--10 keV flux of $\simeq$27\%. Instead fixing the power-law normalization at its 90\% confidence lower limit and re-fitting gives a very similar result: a thermal fraction of $\simeq$26\%. The data therefore seem to imply a 90\% upper limit on the thermal contribution to the flux of $\sim$30\%. With the thermal flux around this upper limit, the fit prefers a lower photon index for the power law, $\simeq$1.3--1.4, similar to the value for XMM-3.


Since none of the other \sw\ spectra have more than 50 counts, we opt to explore this issue further with a simple hardness ratio analysis, rather than try to place fitting constraints on spectra with very few counts. We define our hardness ratio as the net counts in the 3--10 keV band divided by those in the 0.5--3 keV band. This ratio is sensitive to the relative contributions of the two components, since the contribution of the thermal component above 3 keV is small, but a power-law component contributes significantly there. In Figure~\ref{fig:hardness}, we plot this hardness ratio against the (PSF-corrected) net count rate for the four brightest \sw\ observations (filled squares; number of observation indicated next to the symbol). We also use XSPEC to calculate curves based on the {\tt TBnew*(nsatmos+pegpwrlaw)} model and plot them in Figure~\ref{fig:hardness}; these curves depict various combinations of thermal and nonthermal flux in this hardness--intensity diagram (HID). In all cases, we assume the same values for the absorption column, distance, and neutron star mass and radius as in our simultaneous \cxo\ and \xmm\ fit. The bottommost solid curve corresponds to pure thermal spectra; the other three solid curves represent spectra with fixed ratios of thermal and nonthermal flux (i.e., along these curves both the thermal and nonthermal components are varying, but in such a way that the relative contributions of the components to the total unabsorbed 0.5--10 keV flux stays constant) ranging from 25\% power-law contribution (bottom) to 75\% contribution (top). We calculate these four curves for temperatures ranging from that of CXO-11 ($kT^{\infty}_\mathrm{eff}=120.6$ eV) to the maximum temperature for which the {\tt nsatmos} model is valid ($\log(T_\mathrm{eff}/K)=6.5$; $kT^{\infty}_\mathrm{eff}\approx210$ eV); the 75\% curve extends beyond the plot. For the three curves with a power-law contribution, we assume a photon index of $\Gamma=1.7$; this is a compromise between the index seen for XMM-3 (1.40) and the tied index for the other observations in our simultaneous \cxo\ and \xmm\ fit (1.96; we note that the best-fit index value for the penultimate \sw\ observation during outburst at $\sim$$6\times10^{35}\textrm{ erg s}^{-1}$ was also very similar, $1.94\pm0.10$). We also plot curves corresponding to a fixed thermal flux plus a varying power-law flux for two different temperatures: 120.6 eV (as measured for CXO-11; long-dashed curves) and 157.8 eV (as measured for XMM-3; short-dashed curves). In the former case we do this for two photon index values: 1.7 (black curve) and 2 (gray curve); in the latter case for three values: 1.7 (black), 1.4 (gray), and 2 (also gray). The intersections of the solid curves and the $\Gamma=1.7$ dashed curves indicate where along those dashed curves the particular power-law fraction has been reached. Finally, for comparison we also show the locations of the CXO-11 (open circle) and XMM-3 (cross) spectra in this plot. As can be seen from the dashed curves, the value assumed for the photon index has a significant impact on the curves. Changing the photon index would also shift the solid curves (except the purely thermal one) up (lower index) or down (higher index); the higher the power-law fraction, the larger the shift. The location of the XMM-3 data point gives an indication of the magnitude of the shift for the 50\% curve, since the XMM-3 spectrum has a power-law fraction of $\simeq$53\% and a photon index of 1.40.

The HID indicates that \sw\ observations 1, 10, and 11 (especially no.\ 1) are not consistent with the count rate increase being purely due to an increase in the power-law component on top of the underlying baseline quiescent (CXO-11) thermal flux; there has to have been an increase in the thermal flux as well, given the location of these three data points to the right of the long-dashed curves. This is in agreement with the XMM-3 spectrum. However, the location of observation 9 in the diagram indicates that the power-law component dominates in that case, and this is in qualitative agreement with the spectral constraints on the thermal flux in this observation (as discussed above). This implies that the increase in the thermal emission during a flare cannot keep pace with that of the nonthermal emission up to fluxes that high. Observations 9, 10, and 11 are consistent with a flare decay which starts with a decrease in power-law flux ($\Gamma\sim1.5$--2) on top of a fixed thermal component with a temperature similar to that of XMM-3 (thus initially following a trajectory downward with a shape similar to the short-dashed curves); after the power-law fraction reaches $\sim$5\%--25\%, the thermal flux decreases alongside the nonthermal flux (the trajectory then curving to the left and becoming closer to horizontal) until the baseline (CXO-11) temperature has been reached (i.e., the trajectory intersecting with the long-dashed curve). Observation 1, although not part of the same flare, is also consistent with this. The location of XMM-3 perhaps suggests that the photon index of the power-law component in that case (1.40) was lower than in the \sw\ flare; however, it is important to keep in mind that the locations of the two \cxo\ and \xmm\ data points in the HID (as well as the plotted curves, of course) are model dependent, in contrast to the \sw\ data points, and this can affect comparisons. We also emphasize that any discussion of possible trajectories of flares in the HID based on the currently available data is obviously speculative.

\section{Discussion}\label{sec:discussion}

\subsection{Cooling Curve}\label{sec:cooling}

We have presented a \cxo\ observation of the NS-LMXB \src, made in 2010 October, which extends our coverage of the source's current quiescent phase from $\simeq$800 days to $\simeq$1160 days after the end of the 2006--2007 outburst. This new observation suggests that the effective neutron star surface temperature has decreased compared to the preceding \cxo\ observation(s), implying that the neutron star crust may not have reached thermal equilibrium with the core yet. An additional observation---e.g., $\sim$2000 days into quiescence---is needed to conclusively determine whether cooling is still ongoing, and if so, to constrain its rate. The new \cxo\ observation is consistent with the broken power-law fit to the cooling curve presented in \pap. We cannot exclude the possibility that the cooling has followed an exponential decay down to a constant temperature level, although ongoing cooling would make this somewhat unlikely. The simulations of \citet{brown2009} indicate that the form of the cooling curve should approximately be that of a broken power law leveling off to a constant at late times; however, this break is predicted a few hundred days into quiescence and is difficult to reconcile with the much earlier break in our power-law fits (see further discussion of this in \pap). The fits presented in this paper indicate a break in the time range $\sim$25--80 days post-outburst; if the temperature leveled off from the power-law behavior after $\sim$200--300 days (which is somewhat unlikely given the new \cxo\ observation), then the break could conceivably have been a few tens of days later. The range of break times derived here is similar to and consistent with that derived in \pap. The break predicted by \cite{brown2009} is mainly due to a change in the heat capacity of the neutron star crust where the material transitions from a classical to a quantum crystal. A possible alternative explanation for the observed break in the \src\ cooling curve is the existence of a strong nuclear heating source in the crust not considered in the cooling curve simulations; in \pap\ we speculated that the fusion of $^{24}$O \citep{horowitz2008} might be a possible candidate. As discussed in \pap, the rapid initial cooling during the first $\sim$200 days of quiescence indicates a neutron star crust with high thermal conductivity, and possibly suggests low-impurity material \citep{shternin2007,brown2009}. Fitting theoretical models to the cooling curve is necessary to explore in quantitative detail the implications of the observed cooling on the internal properties of the neutron star; this is beyond the scope of this paper. We note again that it is possible that our results are affected to some extent by low-level accretion contributing to the observed thermal emission we use to infer the temperatures; we discuss this in Section~\ref{sec:cooling_effects} and conclude that such effects are unlikely to be large.

It is instructive to compare the observed cooling curve of \src\ to those of the other cooling quasi-persistent transients. The cooling curve of MXB 1659$-$29 can be fitted well with the model of \cite{brown2009}; this fit implies a break at $\sim$300--400 days post-outburst. However, a simple exponential decay to a constant level also provides a good fit to the data \citep{cackett2008}. Both fits indicate that the temperature reached a roughly constant level $\sim$1000--1500 days into quiescence. We note that the implied power-law slopes for MXB 1659$-$29 are much steeper than those of \src. We fitted the temperatures of the first four observations of MXB 1659$-$29 (those made before the cooling leveled off) with a broken power law, using the temperature values given in \cite{cackett2008}; this gave pre- and post-break slopes of $\sim$0.15 and $\sim$0.55 (compared to $\sim$0.03 and $\sim$0.07 for \src). However, the data are very sparse for the early part of quiescence---the first observation was made $\simeq$31--38 days after the end of the outburst and the second observation did not take place until $\sim$400 days into quiescence---and assumptions about the behavior of the source during that period should therefore be regarded with caution. In contrast to MXB 1659$-$29, the cooling curve of KS 1731$-$260 is not well fitted by either the \cite{brown2009} model or an exponential decay, whereas a single unbroken power law provides a good fit, and the source seems to still be slowly cooling more than 3000 days into quiescence; the slope of the best-fit power law is $0.125\pm0.007$ \citep{cackett2010b}.  However, the first observation of KS 1731$-$260 in quiescence did not take place until $\simeq$48--65 days after the end of the outburst, and therefore an early break in the power law similar to the one indicated by the \src\ data cannot be excluded. The four \cxo\ observations of EXO 0748$-$676 made during the first $\sim$600 days of quiescence can be fitted both with an exponential decay to a constant level and with a single unbroken power law (\citealt{degenaar2011a}; the authors also analyze \sw\ observations of the source, which are less constraining but give results consistent with those from the \cxo\ observations). The best-fit power-law slope is $0.03\pm0.01$. An early break cannot be excluded for EXO 0748$-$676 either, since the first \cxo\ observation is only constrained to have taken place sometime during the first $\simeq$15--60 days of quiescence. 

In connection with the discussion above about the different power-law slopes observed for the four sources, it is worthwhile to note that \cite{brown2009} show that the initial slope of the broken power-law curve gives a direct measure of the inward flux near the top of the crust during outburst. They also point out that the inferred early-time slopes for KS 1731$-$260 and MXB 1659$-$29 imply a flux well in excess of that available from electron captures in the outermost layers of the crust. In contrast, the pre-break slope for \src\ (coupled with the mass accretion rate inferred from the observed outburst luminosity) implies a flux which is consistent with the energy available (see \pap). A possible explanation for the apparent inconsistency in the cases of KS 1731$-$260 and MXB 1659$-$29 might be that an early break such as that observed in \src\ took place in those sources as well.

\subsection{Low-level Activity}\label{sec:flaring}

We have also presented results from a \sw\ XRT monitoring campaign of \src\ consisting of 17 short observations made during 2010 April--October. This campaign has clearly established that the increase in flux seen in our third \xmm\ observation was not an isolated event, and that significant temporary elevations of the source flux are repeatedly taking place during quiescence (in the context of \src, we refer to any period outside of extended bright outbursts as quiescence). In 2010 July, we observed a flare that went up to $\simeq$$1\times10^{35}\textrm{ erg s}^{-1}$ and followed a roughly exponential decay from the initial detection with an $e$-folding time of $\sim$3 days (judging from our limited sampling). We note that the luminosity can very possibly have become higher, since the true peak of the flare could easily have been missed due to the rather short duration of the event and the limited sampling. The duration of the flare was at least $\sim$10 days and may have been $\sim$10--20 days longer, but it cannot have started more than two weeks before the initial detection. The first observation of our monitoring program in 2010 April also showed a clearly elevated luminosity of  $\sim$$2\times10^{34}\textrm{ erg s}^{-1}$; this observation may have been made during the decay of a similar flare, although this is clearly highly speculative.

The most natural explanation for these increases in flux is that they are accretion events. Activity at luminosities $\lesssim$$10^{36}\textrm{ erg s}^{-1}$ is for most Galactic transients hard or impossible to detect with all-sky monitors, and dense monitoring of quiescent sources with pointed observations is rare. Our knowledge is therefore limited on how common such low-level activity is in transient sources and what the typical properties of such activity are. A few tens of low-luminosity transients, often referred to as very faint X-ray transients (VFXTs; usually defined as having peak 2--10 keV X-ray luminosities in the range $\sim$$10^{34}$--$10^{36}\textrm{ erg s}^{-1}$), have been identified in the Galaxy (e.g., \citealt{wijnands2006,heinke2009b,degenaar2009a,degenaar2010}). A significant fraction of these have shown type I X-ray bursts and are therefore accreting neutron stars, most likely with low-mass companions (\citealt{degenaar2010} and references therein).  \cite{degenaar2009a,degenaar2010} report on a 4 yr monitoring campaign of the Galactic center with almost daily \sw\ observations for $\sim$9 months of the year. They detected eight faint transients showing a variety of low-level activity (some showed peak luminosities above the VFXT range). Four of these sources showed flares similar to the one we observed from \src, with durations of $\sim$1--2 weeks and peak luminosities in the range $\sim$(0.7--2)$\times10^{35}\textrm{ erg s}^{-1}$; we note that for all four sources only one such short flare was observed during $\sim$4 yr of quasi-daily monitoring, whereas our observations suggest that such events are more frequent in \src. We also note that \cite{degenaar2009a,degenaar2010} quote luminosities in the 2--10 keV band; the peak 2--10 keV luminosity we observed for the \src\ flare---with much sparser sampling, however---is $\sim$$5\times10^{34}\textrm{ erg s}^{-1}$. It has been pointed out that the low long-term average mass accretion rates implied by observations of VFXTs may pose difficulties for binary evolution models in explaining their existence (e.g., \citealt{king2006,degenaar2009a,degenaar2010}). The observed behavior of \src\ supports one simple scenario which could possibly explain this for some sources: if they undergo large outbursts with long recurrence times (perhaps decades or longer) and with sporadic low-level activity in between, then their actual long-term average accretion rates could be much larger than implied by a few years of monitoring. However, this is unlikely to be the case for all such sources, given that none of the known VFXTs have been observed to undergo a large outburst. We also note, as has been pointed out (e.g., \citealt{degenaar2010}), that activity with such low peak luminosities presents a challenge to the disk instability model thought to describe accretion cycles in transient LMXBs \citep{king1998,lasota2001}, and it is not clear that the same mechanism is at work in the flares from \src\ as in the low-luminosity activity of the VFXTs (nor is it clear whether a single mechanism is at work in all VFXTs).


As discussed above and in \pap, the increase in luminosity seen in our third \xmm\ observation resulted from an increase in both thermal and nonthermal flux (and this is probably also the case for the elevations in flux seen with \sw). The physical origin of the nonthermal flux or the extra thermal flux is unclear. A prominent hard (power-law) component is seen in spectra from NS-LMXBs in their low-luminosity hard state; this component is often attributed to some sort of Comptonization (e.g., \citealt{disalvo2002,done2007}). The power-law component we see during flaring may be a lower-luminosity version of the hard state component. It is also unclear whether the variable nonthermal flux seen in all the \cxo\ and \xmm\ spectra has the same origin as the much stronger nonthermal flux seen in XMM-3 and the brighter \sw\ observations. The extra thermal flux observed during flaring may be produced as accreted matter hits the neutron star surface, perhaps in some sort of boundary layer between the surface and an accretion disk (similar to what is seen in NS-LMXBs at higher luminosities). We also note that \cite{zampieri1995} show low-level accretion onto a neutron star surface can under certain conditions result in a spectrum with a thermal (hardened blackbody-like) shape. 

\cite{cackett2010a} show that the thermal flux from the quiescent NS-LMXB Cen X-4 has varied somewhat irregularly over a period of $\simeq$15 yr; the source also has a variable nonthermal component. The origin of this variability is unclear, but the authors argue that the most likely explanation is that the variability in both the thermal and nonthermal components is due to variable low-level accretion. This conclusion is supported by the fact that the two components seem linked together, with their flux ratios staying approximately constant. All the quiescent spectra analyzed by \cite{cackett2010a} show a thermal flux fraction of $\sim$50\%--60\%. This is similar to what we observed in XMM-3, which showed a nonthermal fraction of $\simeq$53\%, but that may well be a coincidence. The other quiescent observations of \src\ have shown power-law fractions ranging from a few percent to $\sim$50\%. As discussed in Section~\ref{sec:swift}, the limited number of counts in our \sw\ spectra do not allow us to place strong constraints on the separate behavior of the thermal and nonthermal components during the flaring we observed. Our simple hardness ratio analysis indicates that some of the increase can be attributed to the thermal component (in agreement with XMM-3), but that at the highest count rates observed the power-law component contributes the majority of the flux. We also note that \cite{cackett2011}, analyzing archival observations of Aql X-1 during quiescence, found a flare with a likely duration of $\sim$60 days and a $\sim$5-fold increase in flux; the data suggest (but are not conclusive) that both the thermal and nonthermal flux rise during the flare.


\subsection{Effects of Low-level Activity on the Equilibrium Temperature}\label{sec:eq_temp}

Although \src\ may still be cooling, the equilibrium surface temperature is probably not very much lower than the current one, given how slow the possible cooling has become at this point. This suggests that the equilibrium thermal luminosity of the source is high compared to other neutron star transients (see, e.g., \citealt{heinke2007,heinke2009a}); by extension, the equilibrium crust and core temperature would then be comparatively high, unless the crust microphysics is considerably different from that used in calculations to date \citep{shternin2007,brown2009}. It is natural to ask whether low-level accretion between large outbursts can to some extent explain the high temperature. To gauge the plausibility of this, we compare an estimate of the long-term average luminosity due to low-level activity to that of bright outbursts; this is equivalent to comparing the long-term average mass accretion rates, assuming that the radiative efficiency of the accretion is the same. To estimate the fluence of the \sw\ flare, we make the conservative assumptions that we caught the flare near its peak and that the pre- and post-peak contributions to the total fluence are of similar size. We also assume that the decay of the flare followed the best-fit exponential mentioned in Section~\ref{sec:swift}. Furthermore, we make the (admittedly highly uncertain) assumption that the typical recurrence time for such flares is 4 months, based on our speculation that the first \sw\ observation was made during the decay of a similar flare. To make a rough bolometric correction, we use the result of \cite{zand2007}, who in the literature on broadband spectra of LMXBs find a typical factor of $2.9\pm1.4$ between reported 0.1--100 keV unabsorbed fluxes and 2--10 keV absorbed fluxes.  For our three brightest \sw\ observations and for XMM-3, we find ratios between the unabsorbed fluxes in the 0.5--10 and 2--10 keV bands in the range $\simeq$2.0--2.2 (using absorbed 2--10 keV fluxes gives a range of $\simeq$2.4--2.7). Since for \src\ the absorption is high, we base our correction on the unabsorbed 2--10 keV flux values we measured and use a correction factor of 1.4 for our 0.5--10 keV luminosities; this number is obviously subject to a large uncertainty, but our overall conclusions are not very sensitive to this. After subtracting the baseline bolometric thermal emission (based on the most recent \cxo\ observation), we find a long-term average flare luminosity of $\sim$$2\times10^{41}\textrm{ erg yr}^{-1}$. Furthermore, assuming that the nonthermal flux we have observed throughout quiescence is due to residual accretion (which may not be the case), and taking an average of the flux values derived from our \cxo\ and \xmm\  observations (excluding XMM-3), we get an average luminosity of $\sim$$7\times10^{40}\textrm{ erg yr}^{-1}$, i.e., roughly a third of the estimated flare contribution. This then gives a total of $\sim$$3\times10^{41}\textrm{ erg yr}^{-1}$ due to low-level activity. The total bolometric energy output of the 2006--2007 outburst is estimated to have been $\sim$$1\times10^{46}\textrm{ erg}$ (see \pap). For the low-level activity to be a significant factor in the total thermal budget of the neutron star would therefore require a recurrence time of at least $\sim$$10^4$ yr for large outbursts like the 2006--2007 one. A recurrence time this long is implausible; assuming that a mass accretion rate $\dot{M}$ results in a bolometric luminosity of $L_\mathrm{bol}=\epsilon\dot{M}c^2$, with $\epsilon=0.2$, this would imply a long-term average mass accretion rate in the system of a few times $10^{-12}\textrm{ }M_\sun\textrm{ yr}^{-1}$, which is hard to reconcile with the high observed surface temperature (see the discussion in \pap\ on the range of recurrence times consistent with standard cooling in the neutron star).

It is interesting to note that of the four transients studied by \cite{degenaar2009a,degenaar2010} that have shown short flares similar to the one in \src, two have also shown longer and more intense activity. For example, GRS 1741.9$-$2853, a confirmed NS-LMXB, showed both a $\sim$1 week flare with estimated peak/average 2--10 keV luminosities of $\sim$$7\times10^{34}$/$3\times10^{34}\textrm{ erg s}^{-1}$ and a $>$13 week outburst with peak/average luminosities of $\sim$$2\times10^{36}$/$1\times10^{36}\textrm{ erg s}^{-1}$, as well as an even brighter 4--5 week outburst with luminosities of $\sim$$1\times10^{37}$/$2\times10^{36}\textrm{ erg s}^{-1}$ (we note again, however, that the $\sim$1 week flare is the only such very short and faint flare that has been detected from the source during four years of monitoring). Based on the detection history of the source in the past decade, \cite{degenaar2010} estimate a recurrence time (not including the $\sim$1 week flare) of $\sim$2 yr. If \src\ experiences similar small outbursts every two years or so with a duration of a few months and an average luminosity of $\sim$$10^{36}\textrm{ erg s}^{-1}$ (any past activity with luminosities $\lesssim$$5\times10^{36}\textrm{ erg s}^{-1}$ would have been hard or impossible to detect with the \xte\ All-Sky Monitor), then a recurrence time of  several hundred years for large outbursts would allow these smaller ones to contribute significantly to the average mass accretion rate. A recurrence time of this order is long but not inconceivable; this would imply a total long-term average mass accretion rate between $10^{-11}$ and $10^{-10}\textrm{ }M_\sun\textrm{ yr}^{-1}$. However, low-level accretion is unlikely to have produced a very large increase in the equilibrium surface temperature, since the fourth-power dependence of the bolometric thermal luminosity on the temperature implies that even if the average low-level accretion rate were of similar size to that from large outbursts, this could not result in more than a $\sim$20\% higher equilibrium temperature than in the absence of any low-level activity. We therefore conclude that it is implausible that low-level accretion can, for example, explain the likely $\sim$2 times higher equilibrium surface temperature of \src\ compared to those of MXB 1659$-$29 and KS 1731$-$260 \citep{cackett2008,cackett2010b}. This difference is more likely mostly due to a higher long-term average accretion rate of long-duration outbursts in \src\ and/or more efficient cooling in the neutron stars in MXB 1659$-$29 and KS~1731$-$260.

\subsection{Effects of Low-level Activity on the Cooling Curve}\label{sec:cooling_effects}

In our discussion, we have assumed that the thermal component in the nonflare quiescent spectra arises entirely from thermal emission from the neutron star surface due to heat from the crust. An important issue that needs to be addressed is whether the thermal component outside flares can to some extent be due to ongoing low-level accretion. A significant contribution from residual accretion to the thermal flux observed would mean that the surface temperatures inferred would not accurately reflect the thermal state of the crust. We see increased thermal emission (along with a rise in nonthermal emission) during flares which are almost certainly accretion events; since we see some nonthermal emission during regular quiescent observations as well, one might suspect that some of the thermal emission observed in those observations might stem from accretion. However, two characteristics of the observed emission indicate that this is unlikely to be a large effect. First, the evolution of the inferred temperature throughout quiescence (excluding the flares) is well behaved, having shown a smooth monotonic decrease (with the exception of CXO-4 and possibly CXO-7, although the latter is consistent with being a statistical fluctuation), with little variation in thermal flux over the last seven observations ($\lesssim$$30\%$). If accretion were in general a significant or even dominant contributor to the thermal emission, much more irregular variability would seem likely; this is what we have observed from the nonthermal emission, which has varied by a factor of $\sim$10 throughout the nonflare observations and by a factor of $\sim$6 in the last seven observations. Second, there does not seem to be any correlation between the thermal and nonthermal flux observed outside the flares, whereas both components rise together during flares. It is hard to see why this behavior would be different outside the flares (i.e., at, presumably, lower accretion rates) compared to during flares if accretion causes a significant part of the thermal emission in both cases. Overall, we conclude that accretion probably does not play a significant role in the nonflare thermal emission. However, we acknowledge that this is a possibility that cannot be conclusively ruled out and could affect our results to some extent.

Another important (and separate) question is whether low-level activity can have a significant effect on the thermal state of the crust and thereby on the thermal surface emission. We first consider the effect---arising from deep crustal heating---of the persistent (but variable) low-level accretion possibly implied by the nonthermal component that has been observed throughout quiescence. We convert the observed average power-law flux to a mass accretion rate as described above, and assume a total heat deposit from nuclear reactions in the neutron star crust of $\simeq$1.9 MeV per accreted nucleon, as calculated by \cite{haensel2008} for a heavy-element envelope on the neutron star composed of $^{56}$Fe nuclear burning ashes. Part of the energy produced will escape the star via neutrino emission, but we can get an upper limit on the photon emission by assuming that all of the energy is radiated away at the neutron star surface. This gives a thermal luminosity contribution of $\sim$$2\times10^{31}\textrm{ erg s}^{-1}$---compared to $\sim$$5\times10^{33}\textrm{ erg s}^{-1}$ currently observed---and corresponds to a completely negligible temperature increase of $\lesssim$$0.1$ eV. We emphasize that the thermal emission we are considering here is due to deep crustal heating and  is not thermal emission instantaneously produced during accretion. Any heating of the neutron star \textit{at the surface} due to accretion should be shallow in extent and thus radiated away very quickly, and would not be expected to have long-term effects on the temperature of the crust.

As estimated in Section~\ref{sec:eq_temp}, the long-term average mass accretion rate due to flares of the sort we observed with \sw\ should only be a factor of $\sim$3 higher than that from the possible persistent low-level accretion (calculated above), and should therefore {\it on average} have a very small effect on the crust temperature. However, in this case, the accretion comes in shorter and more intensive bursts, which could lead to temporary but larger increases in thermal flux from the surface. Calculating the effect of short-duration accretion events on the surface temperature is complicated by many factors. The nuclear reactions which contribute to the total heat deposit per nucleon take place at various depths in the crust and the thermal diffusion time to the surface varies greatly, from days to many hundreds of days \citep{brown2009}. The strongest heat sources, producing the majority of the total heat deposit, are at densities of $10^{12}$--$10^{13}\textrm{ g cm}^{-3}$ \citep{haensel2008}, at depths from which the thermal diffusion time to the surface is likely hundreds of days \citep{brown2009}. Diffusion will also cause the effect of any individual reaction on the surface temperature to be spread out in time. In addition, part of the heat will flow into the core rather than to the surface; some fraction will be radiated away through neutrinos. Making the same assumptions as before about the fluence of the flare we observed, we estimate the total mass accreted during the flare to be $\sim$$4\times10^{20}$ g; this gives a total heat deposit of $\sim$$7\times10^{38}$ erg in the crust if we again assume that each nucleon contributes $\simeq$1.9 MeV. We can make a very crude order-of-magnitude estimate of the upper limit to the luminosity increase due to the flare by assuming that all this heat is radiated at the surface on the typical thermal diffusion timescale of the crust, $\sim$$10^2$ days \citep{brown2009}. This gives a luminosity of $\sim$$10^{32}\textrm{ erg s}^{-1}$, which would imply a temperature increase of $\lesssim$0.5 eV---virtually undetectable with current instruments. We note, however, that a total amount of accreted mass that is one or two orders of magnitude higher---conceivable for longer and more intense low-level activity---would place this upper limit high enough for a detectable effect on the temperature.

\cite{ushomirsky2001} study the time-variable quiescent luminosity of a neutron star undergoing very short (1 day) accretion outbursts with recurrence times of 1 or 30 yr, and find that this luminosity depends sensitively on the microphysics of the crust and core. They consider both standard and rapid core cooling, as well as both high and low thermal conductivity for the crust. In their simulations, even a total accreted outburst mass three orders of magnitude higher than what we estimate for the observed flare in \src\ does not result in a luminosity increase---due to crustal heating---of more than a few times $10^{32}\textrm{ erg s}^{-1}$; this holds regardless of what assumptions they make for the core cooling and crust conductivity. This suggests that periods of low-level activity that are considerably brighter and longer than what we have observed in \src\ would not be able to have an appreciable effect on the surface temperature. We stress, however, that there is considerable uncertainty in the microphysics of the neutron star crust and core, and we cannot exclude the possibility that more refined calculations---and calculations more tailored to the observed behavior of \src---would lead to different conclusions. 



The fourth \cxo\ observation indicates increased thermal flux without an associated rise in nonthermal flux, and the preceding observation (XMM-3) strongly indicates ongoing accretion. It is therefore natural to ask whether the possibly elevated surface temperature observed in CXO-4 can be attributed to crustal heating from the presumed XMM-3 accretion event. We conclude that this is unlikely given the results discussed above and how close in time XMM-3 and CXO-4 were. These observations were only separated by $\simeq$73 days and any enhanced activity cannot have started more than $\sim$120 days before CXO-4, whereas most of the crustal heating takes place at depths from which the thermal diffusion time to the surface is thought to be hundreds of days, as mentioned above.



\section{Summary}\label{sec:summary}

We have presented a new \cxo\ observation, made in 2010 October, of the transient NS-LMXB \src, which entered quiescence in 2007 August after an extraordinarily luminous 19 month outburst. This observation extends our monitoring of the source in quiescence from $\simeq$800 days to $\simeq$1160 days since the end of the outburst. We have also presented the results of a \sw\ monitoring program of \src, which took place during 2010 April--October, and whose purpose was to investigate possible low-level activity from the source.

The new \cxo\ observation indicates that the effective surface temperature of the neutron star may have decreased since the preceding \cxo\ observation(s) in 2009, implying that the neutron star crust may still be slowly cooling toward thermal equilibrium with the neutron star core after having been heated during the 2006--2007 outburst. An additional observation further into quiescence is needed to conclusively determine whether cooling is still ongoing, and if so, to constrain its rate. \citet{brown2009} calculate that the form of the cooling curve should be a broken power law leveling off to a constant at late times. With the present data, the shape of the overall cooling curve is consistent with a broken power law. However, the observed break (at $\sim$25--80 days into quiescence) is much earlier than the break predicted by theory (at a few hundred days post-outburst) and may therefore have a different origin. We can also not exclude that the cooling has followed an exponential decay to a constant level, although ongoing cooling would make this somewhat unlikely. The rapid initial cooling during the first $\sim$200 days of quiescence strongly indicates a highly conductive neutron star crust, and possibly suggests low-impurity material \citep{shternin2007,brown2009}. The high current surface temperature (which corresponds to a bolometric thermal luminosity of $\sim$$5\times10^{33}\textrm{ erg s}^{-1}$), coupled with the slow (or possibly nonexistent) current cooling, suggests that the equilibrium crust/core temperature of \src\ is high compared to other transient NS-LMXBs, unless the microphysics of the crust is considerably different from that used in calculations to date \citep{shternin2007,brown2009}.

We observed a large temporary increase in luminosity with \xmm\ $\simeq$230 days into quiescence. This prompted us to undertake a 7 month \sw\ monitoring program of \src\ with short observations once every two weeks. During this time, we observed a short-term ($\sim$10--20 day) flare---presumably arising from low-level accretion---which went up to a luminosity of at least $\sim$$1\times10^{35}\textrm{ erg s}^{-1}$, i.e., $\sim$20 times higher than the normal quiescent level. We also observed a smaller increase in luminosity at the beginning of the program. Accretion during events like these will lead to some heating in the crust. A simple analysis suggests that flares of the magnitude observed are not likely to have significantly affected the equilibrium crust/core temperature of the neutron star and should not be able to have a significant impact on the cooling curve. However, it is possible that brighter and longer periods of low-level activity---as have been observed in some faint Galactic neutron star transients---may have had an appreciable effect on the equilibrium temperature of the neutron star. It is important that the source be monitored further to better constrain its low-level activity in quiescence.

\acknowledgements

This work was supported by \cxo\ Award GO9-0057X and \sw\ Award NNX10AK87G. We thank the \sw\ Mission Operations Center for scheduling the DDT observations and the referee for helpful comments which served to improve the paper. This research has made use of data obtained from the High Energy Astrophysics Science Archive Research Center (HEASARC), provided by NASA's Goddard Space Flight Center.

\bibliography{references}

\end{document}